\documentclass[aps,prl,twocolumn,nofootinbib,superscriptaddress,floatfix]{revtex4-1}  
\usepackage{graphicx}  
\usepackage{amsmath}
\usepackage{amssymb}   
\usepackage{color}
\usepackage{hyperref}
\usepackage{ulem}

\newcommand{\fnl}{f_{\rm NL}}

\newcommand{\me}{\mathrm{e}}

\newcommand{\de}{\partial}

\newcommand{\mpl}{M_{\rm Pl}}

\begin{document}

\title{Mild quasi-local non-Gaussianity as a signature of modified gravity during inflation}

\author{Nicola Bartolo}
  \email{nicola.bartolo@pd.infn.it}
  \affiliation{Dipartimento di Fisica e Astronomia G. Galilei, Universit\`a degli Studi di Padova, I-35131 Padova, Italy.}
  \affiliation{INFN, Sezione di Padova, I-35131 Padova, Italy.}

\author{Dario Cannone}
  \email{dario.cannone@pd.infn.it}
  \affiliation{Dipartimento di Fisica e Astronomia G. Galilei, Universit\`a degli Studi di Padova, I-35131 Padova, Italy.}
  \affiliation{INFN, Sezione di Padova, I-35131 Padova, Italy.}

\author{Raul Jimenez}
  \email{raul.jimenez@icc.ub.edu}
 \affiliation{ICREA \& ICC, University of Barcelona, Marti i Franques 1, Barcelona 08028, Spain.} 
 \affiliation{Institute for Applied Computational Science, Harvard University, MA 02138, USA.}

\author{Sabino Matarrese}
  \email{sabino.matarrese@pd.infn.it}
   \affiliation{Dipartimento di Fisica e Astronomia G. Galilei, Universit\`a degli Studi di Padova, I-35131 Padova, Italy.}
  \affiliation{INFN, Sezione di Padova, I-35131 Padova, Italy.}
  \affiliation{Gran Sasso Science Institute (INFN), viale F. Crispi 7, 67100 L'Aquila, Italy.}

\author{Licia Verde}
  \email{liciaverde@icc.ub.edu}
  \affiliation{ICREA (Instituci\'o Catalana de Recerca i Estudis Avan\c{c}at) \& ICC, University of Barcelona, Marti i Franques 1, Barcelona 08028, Spain} 
  \affiliation{Institute of Theoretical Astrophysics, University of Oslo, 0315 Oslo, Norway. \\}

\date{\today}

\begin{abstract}
We show that modifications of Einstein gravity during inflation could leave 
potentially measurable imprints on cosmological
observables in the form of non-Gaussian perturbations.  This is due to the fact that these modifications
appear in the form of an extra field that could have non-trivial interactions with the inflaton.
We show it explicitly for the case $R+\alpha R^2$, where  nearly scale-invariant non-Gaussianity  at the level
of $f_{\rm NL} \approx -(1$ to $30)$ can be obtained,  in a quasi-local  configuration.
\end{abstract}

\pacs{}
\maketitle

The current inflationary paradigm \cite{Brout:1977ix,Starobinsky:1980te,Kazanas:1980tx,Mukhanov:1981xt,Guth:1980zm,Sato:1980yn,Linde:1981mu} is the most economical at successfully describing many observed features in the Universe, from its homogeneity, flatness and size, to the origin of the structure in the Universe as quantum fluctuations e.g.,~ \citep{Ade:2013uln,Ade:2013ydc}.
In the vast majority of inflationary models, Einstein gravity is assumed as the correct description of gravity. 
However, it might be that Einstein gravity is not the correct description of gravity at very high energies either via a true
modification of General Relativity or because quantum effects become relevant. Departures from Einstein gravity during inflation  
have been considered in the first inflationary model proposed \cite{Starobinsky}, in Ref.~\cite{Berkin:1991nm,Mollerach,Starobinsky:2001xq,DiMarco:2002eb,Maldacena:2011nz,Soda:2011am,Shiraishi:2011st} and most recently  in Ref.~\cite{Xia}. 
In Ref.~\cite{Maldacena:2011nz} (see also~\cite{Soda:2011am,Shiraishi:2011st})
graviton non-Gaussianities are considered beyond ordinary Einstein gravity. However such non-Gaussianities are well below the sensitivity of future measurements and in fact well below the cosmic variance limit for the full sky.
In this Letter we investigate if  deviations from General Relativity (GR) could be observable and measurable in the sky through the enhancement of non-Gaussianity (NG) of curvature perturbations. In the simplest models of inflation with standard gravity~\footnote{
Or inflation models within modified gravity which can be described as General Relativity (GR) plus single-field slow-roll inflation.},  the amount of primordial non-Gaussianity (NG)  is too small to be measurable,  the NG parameter  $f_{\rm NL}$ being  $\sim\mathcal{O}(\epsilon)$ \cite{Gangui:1993tt,Acquaviva2002, Maldacena2002}.

NG has been recognised as a powerful tool to learn about fundamental physics at play during inflation, being  a  probe of the interactions of the field(s) driving inflation. Other statistics, such as the power spectrum, do not carry  as {\it specific} signatures as NG does. For this reason we expect that the effect of modifying gravity will leave specific  signatures on the departures from Gaussianity.
We find that this is the case, in particular we show that modifications
of Einstein gravity, if already relevant during the epoch of inflation, could lead to a measurable non-Gaussian
signature in the cosmological fluctuation field. Such non-Gaussian signatures would be the imprints of  
departures from GR that, on the other hand, might be much harder to probe in the power spectrum of scalar perturbations. Also, 
we will show that, for a large part of the parameter space, the generated non-Gaussianities have a quasi-local shape. This is observationally promising given that future LSS surveys can be sensitive to values of local NG 
$f_{\rm NL} \sim \mathcal{O}(1)$ or even smaller~(see, e.g., \cite{Carbone:2008iz,Verde:2009hy,Giannantonio:2011ya,SeljakCV}).

\begin{figure*}
\hspace*{-6mm}\includegraphics[width=2.1\columnwidth]{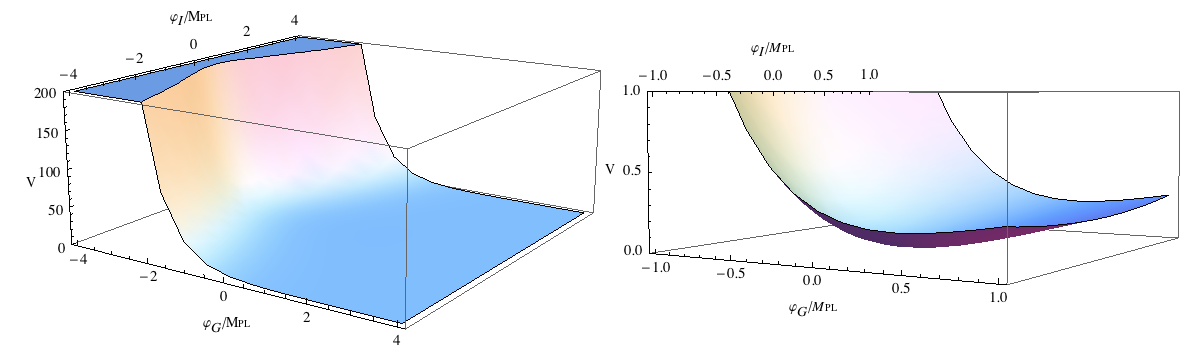}
\caption{Potential as a function of the two scalar fields. $\varphi_G$ describes the ``scalaron'' field that accounts for
modifications of Einstein gravity while $\varphi_I$ is the one driving inflation. Significant non-Gaussianities ($|f_{\rm NL}| \approx 1-30$) are generated for generic initial field values, provided $\varphi_G > -3$. Parameters are chosen for illustration purposes. In particular we chose a quadratic potential \cite{1985PhLB..163..331P} for the inflaton field $\varphi_I$. The right panel shows the potential around the minimum.}
\label{fig:potential}
\end{figure*}

Let us start from a Lagrangian that contains all generally covariant terms  up to two derivatives built with the metric and one scalar field,
that we will assume to drive inflation \cite{Weinberg2008}:
\begin{eqnarray} \label{L_general}
L&=&\sqrt{-g}\Bigg[\frac{1}{2}M_{Pl}^2\Omega(\psi)^2R-\frac{1}{2}h(\psi)g^{\mu\nu}\de_\mu\psi\de^\mu\psi-U(\psi)\nonumber\\&&
+f_1(\psi)\Big(g^{\mu\nu}\de_\mu\psi\de_\nu\psi\Big)^2
+f_2(\psi)g^{\rho\sigma}\de_\rho\psi\de_\sigma\psi\Box\psi\nonumber\\&&
+f_3(\psi)\Big(\Box \psi\Big)^2
+f_4(\psi)R^{\mu\nu}\de_\mu\psi\de_\nu\psi \nonumber\\&&
+f_5(\psi)R\, g^{\mu\nu}\de_\mu\psi\de_\nu\psi\nonumber
+f_6(\psi)R\, \Box \psi
+f_7(\psi)R^2\\&&
+f_8(\psi)R^{\mu\nu}R_{\mu\nu}
+f_9(\psi)C^{\mu\nu\rho\sigma}C_{\mu\nu\rho\sigma}\Bigg]
\nonumber\\
&&+f_{10}(\psi)\epsilon^{\mu\nu\rho\sigma}C_{\mu\nu}{}^{\kappa\lambda}C_{\rho\sigma\kappa\lambda}\,.
\end{eqnarray}
If the inflaton $\psi$ is slowly-rolling, then the functions $\Omega(\psi)$, $h(\psi)$ and $f_i(\psi)$ are
varying slowly and can be simply treated as constants
up to slow-roll corrections, which we will neglect. In this case, the Weyl-squared terms can be recast as a surface term (the Gauss-Bonnet term)  plus
$R^2$ and $R_{\mu\nu}R^{\mu\nu}$, which can then be reabsorbed. Moreover, in order to avoid ghosts, the terms
proportional to  $f_2$, $f_3$, $f_6$ and $f_8$ will be set to zero, as well as $f_{10}$ as we are not interested in parity
violating signatures. We are interested only in the terms that could give rise to a possibly enhanced
local (or quasi-local) NG in the squeezed limit, different from  the well-known result $\fnl\sim\mathcal{O}(\epsilon)$ that is valid in
standard gravity \cite{Gangui:1993tt,Acquaviva2002, Maldacena2002}. Therefore we will not consider inflaton derivative self-interactions, which are known to generate
NG mainly in the equilateral configuration. This is valid also for the ghost-free combination that can be built
with the operators proportional to $f_4$ and $f_5$ \cite{Germani2011}, which would not generate significant NG in the
local configuration. The only term left to consider is therefore the term $R^2$, which is nothing else than the first term
in an expansion in powers of the Ricci scalar of a more general $f(R)$ theory:
\begin{equation} \label{f(R)}
 \mathcal{L}=\sqrt{-g}\left[f(R)-\frac{1}{2}g^{\mu\nu}\de_\mu\psi\de_\nu\psi-U(\psi)\right] \; .
\end{equation}
This action describes one more degree of freedom associated to the $f(R)$ term.
Through a standard procedure we use an auxiliary field $f'(\chi)=\mpl^2 \phi/2$ to recast the action in the form
\begin{equation}
 \mathcal{L}=\sqrt{-g}\left[\frac{1}{2}\mpl^2\phi R+\Lambda(\phi)-\frac{1}{2}g^{\mu\nu}\de_\mu\psi\de_\nu\psi-U(\psi)\right] \; ,
\end{equation}
where $\Lambda(\phi) =f(\chi(\phi))-\mpl^2 \phi \chi/2$.

By performing a Weyl transformation $g_{\mu\nu}\to\me^{-2\omega}g_{\mu\nu}$, with $\me^{2\omega}=\phi$, to go to the Einstein frame,
 the action appears as a two-field interacting model:
 \begin{equation} \label{L_2field}
 \begin{array}{lcl}
  \tilde{\mathcal{L}} & = \sqrt{-g}\Bigg[&\!\displaystyle\frac{1}{2}\mpl^2R-\frac{1}{2}g^{\mu\nu}\gamma_{ab}\de_\mu\varphi^a\de_\nu\varphi^b \\
		      &   &\displaystyle -U_1(\varphi_1)-\me^{-4\varphi_1/\sqrt{6}\mpl}U(\varphi_2)\Bigg] \; ,
 \end{array}
 \end{equation}
 where $a,b = 1,2$ we have normalized the fields as
 \begin{equation}
  \sqrt{6}\mpl\omega=\varphi_1 \;,\qquad\psi=\varphi_2 \;,
 \end{equation}
 defined $U_1$ as
 \begin{equation}
  U_1(\varphi_1)=-\me^{-4\varphi_1/\sqrt{6}\mpl}\Lambda\left(\phi\left(\omega\left(\varphi_1\right)\right)\right) \, ,
 \end{equation}
 and defined the field metric
 \begin{equation}
  \gamma_{ab}=\left(\begin{array}{cc}
                     1 & 0 \\
                     0 & \me^{-2\varphi_1/\sqrt{6}\mpl}
                    \end{array}\right)\,.
 \end{equation}
As expected, there is an equivalence between ``$f(R)$+scalar'' and a two-field model with a specific field metric,
a generic potential for $\varphi_1$ and a ``conformally-stretched'' potential for $\varphi_2$. Then it is conceivable that
the interactions between the two fields could induce some observable effects, possibly enhancing also local NG
to an observable level. It is important to note here that if both  fields contribute to the dynamics of the background,
we should rigorously impose slow-roll conditions on both of them. However, if the field associated to the $R^2$ terms is
subdominant, then this condition could be relaxed and its possible NG could be transferred to the inflaton field.
In the Einstein frame this is equivalent to a transfer of non-Gaussian isocurvature perturbations to the adiabatic perturbation mode~\cite{Bartolo:2001cw}.
To study this effect, we will consider $f(R)=\frac{1}{2}\mpl^2 R+R^2/12M^2$.

\begin{figure}
\hspace*{-6mm}\includegraphics[width=\columnwidth]{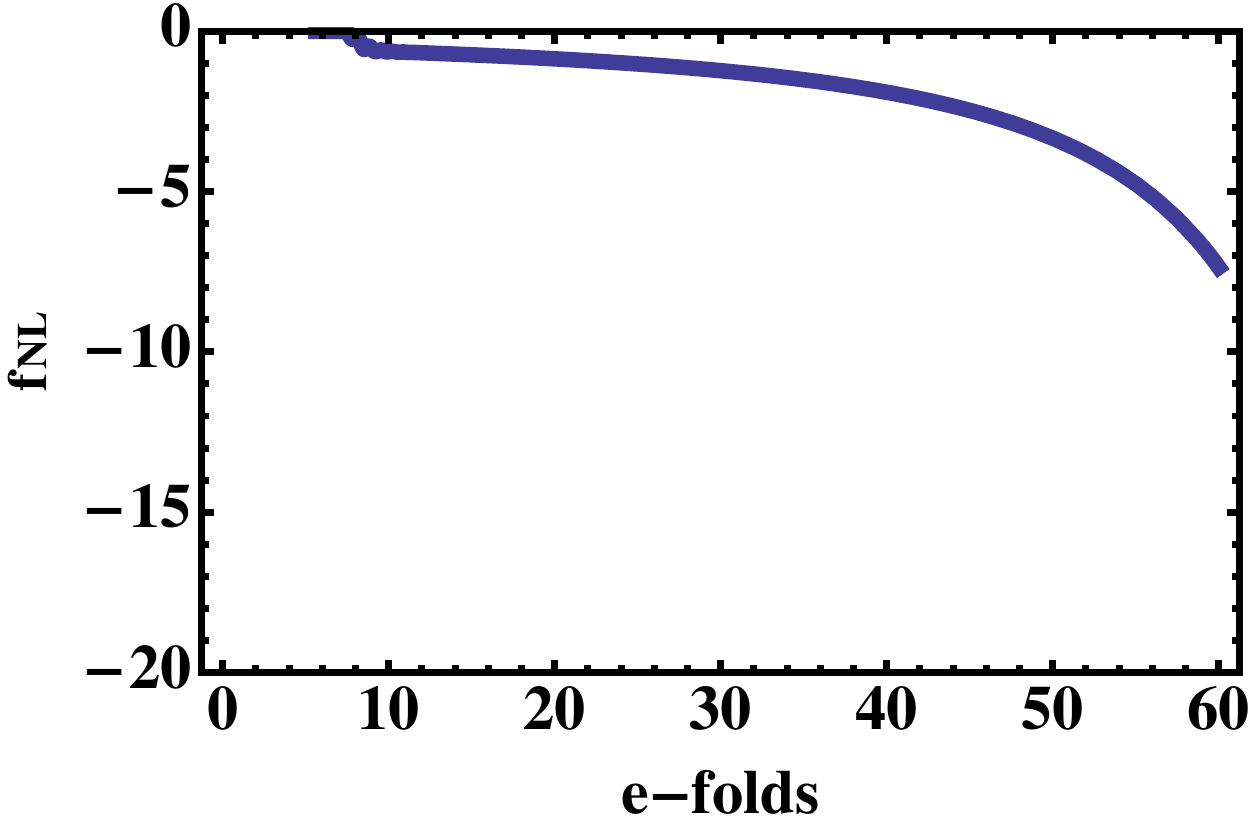}
\caption{The NG parameter $f_{\rm NL}$ as a function of number of e-folds for $\alpha(\nu)=1$, $M=10^{-3}$ and $m=10^{-8/3}$ in units of $M_{\rm Pl}$ to illustrate the scale dependence;  $f_{\rm NL}$ can be smaller than $-1$ for fairly generic conditions.}
\label{fig:fnl}
\end{figure}

This choice is motivated by the fact that  it corresponds to the
leading order term
in an expansion of a generic $f(R)$ in powers of
$R$ (or equivalently in derivatives of the metric). In this case, we obtain a complete potential $V(\varphi_1,\varphi_2)$ given by:
\begin{equation}\label{potential}
 \begin{array}{lcl}
  V(\varphi_1,\varphi_2) & = &\displaystyle\frac{3}{4}M^2\mpl^4\left(1-\me^{-2\varphi_1/\sqrt{6}\mpl}\right)^2\\
                         &   &\displaystyle +\me^{-4\varphi_1/\sqrt{6}\mpl}U(\varphi_2) \; .
 \end{array}
\end{equation}
It is clear that if the field $\varphi_1$ is very heavy and the scale of the new physics induced by the $R^2$ term is
much higher than the energy scale of the inflaton $\varphi_2$, then its effect should be vanishingly small. Indeed, if $\varphi_1$ is heavy enough, it could not be excited during inflation and its kinetic energy would be completely negligible. Therefore we could
integrate it out of the action \eqref{L_2field}, coming back to a standard effective single-field scenario.  This would correspond
to a value of $M\sim1$ or higher, which implies that the new physics simply enters at the Planck scale or beyond.
On the other hand, lowering the scale
$M\lesssim1$, the first regime we encounter is the quasi-single field regime \cite{Chen2009}. Progressively reducing  the  value of $M$, other regimes are possible:  first the multi-field inflation where both scalar fields  are actively at play and then, when the field $\varphi_1$ dominates the dynamics, 
single-field Starobinsky inflation \cite{Starobinsky}. Hereafter, we adopt a monomial potential $U(\varphi_2) = m^{4-\beta} \varphi_2^{\beta}$, with $\beta < 2$ (motivated by current Planck-satellite constraints). Our results are insensitive to the choice of $\beta$.

We are interested in the quasi single-field regime, as observables do not depend on the particular choice of the initial conditions. In this sense we look for {\it generic} predictions.
In this case, assuming that
the adiabatic direction is given by $\varphi_2\equiv \varphi_I$, we  obtain non-trivial effects from the coupling with the isocurvature
field $\varphi_1\equiv \varphi_G$. (Here  by using the subscripts I and G we have made explicit that the field $\varphi_I$ is the inflation  and $\varphi_G$ describes the modifications of gravity). To make an estimate of the magnitude of the effect, we can expand the action  Eq. ~\eqref{L_2field} in the flat gauge and ignore metric perturbations for
simplicity. At second order, we find the leading transfer vertex
\begin{equation}
\label{int1}
 \delta \mathcal{L}_2 = \frac{2}{\sqrt{6} \mpl} \me^{\frac{-2 \bar\varphi_G}{\sqrt{6} \mpl}}\dot{\bar\varphi}_I \delta \varphi_G \delta \dot \varphi_I \;,
\end{equation}
where the bar refers to homogeneous quantities computed on the background. At third order, as the isocurvature potential $U_1'''$
is not subject to slow-roll conditions, the leading vertex is
\begin{equation}
\label{int2}
 \delta\mathcal{L}_3 = -\frac{1}{6}U_1'''(\bar\varphi_I)\delta\varphi_G^3.
\end{equation}
Therefore we expect a contribution to the bispectrum of  size
\begin{eqnarray}
\label{eq:11}
 \fnl & \simeq & \displaystyle\alpha(\nu)\,\left(\widehat{\delta\mathcal{L}}_2\right)^3\, 
\widehat{\delta \mathcal{L}}_3\,\mathcal{P}_\zeta^{-1/2} \\ \nonumber
      &   =    & - \displaystyle\frac{4}{9 \pi}\alpha(\nu)\,  \frac{\mathcal{P}_\zeta^{-1}}
{\sqrt{\epsilon}}\, M^2 \left[ \epsilon-3 \left( \frac{\dot{M}_{\rm Pl,eff}}{H M_{\rm Pl,eff}} 
\right)^2 \right]^{3/2} \\ \nonumber
      &  \times  &        \displaystyle\left[   \left(   \frac{M_{\rm Pl,eff}}{M_{\rm 
Pl}}\right)^2-4 \right] \left( \frac{M_{\rm Pl,eff}}{M_{\rm Pl}}\right)^{-7}
\end{eqnarray}
 where $\widehat{\delta\mathcal{L}}_2$ and $\widehat{\delta \mathcal{L}}_3$ are the vertices of the interaction terms, Eqs.~(\ref{int1}-\ref{int2}), $\nu=\sqrt{9/4-(M_{\rm eff}/H)^2}$, $M_{\rm eff}$ is the effective mass of the isocurvature mode
 and $\epsilon$ the total slow-roll parameter. In Eq.~(\ref{eq:11}) $M_{\rm Pl,eff}= M_{\rm Pl}\,\, \me^{\varphi_G/\sqrt{6} M_{\rm Pl}}$ is the effective (reduced) Planck mass during inflation in the Jordan frame. The numerical factor $\alpha(\nu)$ can range from $0.2$, for heavier isocurvatons, to approximately $300$; however,  in the perturbative regime, NG can gain at most an effective enhancement factor proportional to the number of e-foldings, see \cite{Chen2009}.

The shape of the potential as a function of the two fields $\varphi_I$ and $\varphi_G$ is shown in Fig.~\ref{fig:potential}. On the left panel one can appreciate that the $\varphi_I$ direction is flat but there are values of $\varphi_G$  where the potential is steep. On the right panel we show the region around the global minimum.  Figure (\ref{fig:fnl}) shows the NG parameter $f_{\rm NL}$ as a function of e-folds adopting $U(\varphi_I) = m^3 \varphi$; our results are not sensitive to the specific value adopted for $\beta$.
As an example, for $M= 10^{-3}$ and $m=10^{-8/3}$, in Planck units, we obtain $\fnl\sim\mathcal{O}(-3)$, for initial values of the field  $\varphi_G = 3, \varphi_I = 12$. Note the nearly scale invariant dependence. For this particular example at 60 e-folds the field abandons slow-roll and re-heating starts.
The characteristic shape of this kind of NG is intermediate between an equilateral shape, which is reached for small
values of $\nu$ i.e., towards a single-field regime, and a local shape, for $\nu\geq1/2$ i.e., closer to a multi-field scenario. 

In this set up $f_{\rm NL}$ is generically negative. A quasi-local shape with $f_{\rm NL}\approx -1$ to $-30$ can thus be  achieved without necessity of much fine tuning. The value of $f_{\rm NL}$ scales as a function of the masses of the two potentials,
\begin{equation}
f_{\rm NL} \propto - (M M_{\rm Pl}/m)^2 \alpha(\nu).
\end{equation}
This makes it possible to test deviations from GR, including quantum corrections of Einstein gravity, a couple of
orders of magnitude above the mass scale of the inflaton.
Note that   Eq.~(\ref{eq:11})  gives a ``consistency relation" between the amplitude of NG and its shape. In fact, $f_{\rm NL}$ measures departures from the effective gravitational constant $G_{\rm eff}$ during inflation as $G_{\rm eff}/G_{\rm GR} = \me^{-\varphi_G/\sqrt{6} M_{\rm Pl}}$.

To summarise, we have explored whether signatures of modified gravity during the period of inflation can produce observable effects.
To be used  to gain insight into the physics at play during inflation, these effects  should be specific and not easily  mimicked by standard gravity, yet arising under fairly generic conditions. For this reason we concentrated on local (or quasi-local) NG:  departures from Gaussianity are ${\cal O} (\epsilon)$  in standard gravity single-field inflation (and higher-derivative inflaton self-interactions generate equilateral-like NG, the same being true for various gravity theories with one scalar degree of freedom that can be described in terms of a Horndeski theory, such as Galileon models~\cite{Burrage:2010cu} -- for a summary of predictions see, e.g.,~\cite{Tsujikawa:2013ila}).
Large non-Gaussianities can arise in multi-field inflation but also other observational signatures can be generated such as isocurvature modes and breaking the tensor consistency relation.
We have found that it is possible, in a very generic set-up, for modifications of gravity to 
generate deviations from Gaussian initial conditions where the NG is  close to the local type and has values  $f_{\rm NL} \approx -1\,\,{\rm to } -30$. 

 It is interesting to note that in the same way that gravity, via its relativistic corrections, enhances the level of NG to $\fnl\sim\mathcal{O}(-1)$ right after inflation (as pioneered by \cite{Bartolo:2005xa,Verde:2009hy}), a modification of GR {\it during} inflation will lead to an enhancement of similar magnitude.

 For quasi-local shapes  NG is near maximal in the squeezed limit and the squeezed limit is made observationally accessible in the so-called large-scale halo bias.

Thanks to the halo bias effect, a local NG of this amplitude  is expected to be  measurable  in forthcoming and future LSS surveys~(see, e.g., \cite{Dalaletal:2008, MV08, Carbone:2008iz,Verde:2009hy,Giannantonio:2011ya})  if systematic effects can be kept under control (e.g., \cite{Leistedt}). On the other hand, the  departures from exact single-field behaviour leave some imprint on the shape of NG, and in particular on the squeezed-limit dependence of the bispectrum on the (small) momentum.  In fact, since the shape of the effective potential, Eq.~(\ref{potential}), is given,  there is a ``consistency relation" linking the amplitude of non-Gaussianity, $f_{\rm NL}$, to its shape (i.e., the parameter $\nu$).
 For large enough values of $f_{\rm NL}$ it is possible to constrain the scale-dependence  of the bispectrum in the squeezed limit and hence $\nu$, from forthcoming surveys \cite{Norenaetal12,SefusattiChen12}. 
Thus, in case of a detection of NG,  it may be possible to test the ``consistency relation" between  amplitude and shape. If such consistency relation were found to be satisfied to sufficient precision,  it would require a fine tuning to be produced by any multi/quasi-single field inflation. Conversely, it is a fairly generic prediction  of GR modification effects at high energies.

Further, because the non-inflating field is related to gravity, the ratio between $r$ (the tensor-to-scalar ratio) and its power law slope ($n_T$) will be modified from the standard single field relation---with its counterpart in the two-field description in the Einstein frame~\cite{Bartolo:2001rt,Wands:2002bn}. A given form for $f(R)$ (corresponding to a given shape of $U_1(\varphi_G)$) will break the standard consistency relation in a specific way. 

Notice also that a specific running of the NG parameter $f_{\rm NL}$ in Eq.~(\ref{eq:11}) can be left imprinted by the dynamics of the ``scalaron'' field $\varphi_G$, and interestingly the NG running will be correlated with the running of the scalar spectral index~\cite{Chen2009}. Specific signatures in the trispectrum of curvature perturbations, similar to those featured  in Eq.~(\ref{eq:11}), arise as well.

To conclude,  these  findings, if supported by  data, would yield clear  insights into the physical mechanism behind inflation. Conversely, a null result would place limits on possible
departures from GR at the energy scale of inflation, 20 orders of
magnitude beyond what has been currently tested. 

\mbox{}

\paragraph{Acknowledgments.---}
The work of NB and SM was partially
supported by the ASI/INAF Agreement 2014-024-R.0 for the 
Planck LFI Activity of Phase E2. DC thanks the ICG Portsmouth for its warm hospitality during the last stages of this work. LV is supported by the European Research Council under the European Community's Seventh Framework Programme FP7-IDEAS-Phys.LSS 240117. LV and RJ acknowledge Mineco grant FPA2011-29678- C02-02.  NB and SM thank Frederico Arroja and Maresuke Shiraishi for useful discussions.


%

\end{document}